\documentclass[conference]{IEEEtran}

\usepackage{graphicx}
\usepackage{amsmath}
\usepackage{url}
\usepackage{amsfonts}
\usepackage{subfig}
\usepackage[linesnumbered,ruled]{algorithm2e}
\usepackage{verbatim}
\usepackage[noend]{algpseudocode}
\usepackage{xspace}
\usepackage{cite}

\usepackage[T1]{fontenc}
\usepackage{inconsolata}

\usepackage{amsmath,amsthm,amssymb,multirow,tabls,bigstrut}
\usepackage{tabularx}
\usepackage[table,xcdraw]{xcolor}
\usepackage{diagbox}
\usepackage[left=0.6in, right=0.6in, top=0.7in, bottom=1in]{geometry}
\usepackage{color}

\newcommand{\showComments}{yes}

\newtheorem{definition}{\bf Definition}

\def\squareforqed{\hbox{\rlap{$\sqcap$}$\sqcup$}}
\def\qed{\ifmmode\squareforqed\else{\unskip\nobreak\hfil
\penalty50\hskip1em\null\nobreak\hfil\squareforqed
\parfillskip=0pt\finalhyphendemerits=0\endgraf}\fi}

\definecolor{placeholderbg}{rgb}{0.85,0.85,0.85}

\newcommand{\note}[2]{
    \ifthenelse{\equal{\showComments}{yes}}{\textcolor{#1}{#2}}{}
}

\title{LERC: Coordinated Cache Management\\for Data-Parallel Systems}
\author{
\IEEEauthorblockN{Yinghao Yu$^\dagger$, Wei Wang$^\dagger$, Jun Zhang$^\dagger$, Khaled B. Letaief$^{\dagger\ddagger}$\\}
\IEEEauthorblockA{$^\dagger$Hong Kong University of Science and Technology\\
$^\ddagger$Hamad bin Khalifa University, Doha, Qatar\\
\tt $^\dagger$\{yyuau, weiwa, eejzhang, eekhaled\}@ust.hk $^\ddagger$kletaief@hbku.edu.qa
}
\thanks{This work was supported by the Hong Kong Research Grants Council under Grant No. 16200214.}
}

\graphicspath{{figures/}}
\usepackage{microtype}  

\IEEEoverridecommandlockouts
\begin{document}

 \maketitle

\begin{abstract}

Memory caches are being aggressively used in today's data-parallel frameworks
such as Spark, Tez and Storm. By caching input and intermediate data in memory, compute tasks can witness speedup by orders of magnitude. To maximize the chance of
in-memory data access, existing cache algorithms, be it recency- or
frequency-based, settle on \emph{cache hit ratio} as the optimization
objective. However, unlike the conventional belief, we show in this paper
that simply pursuing a higher cache hit ratio of individual data blocks does
not necessarily translate into faster task completion in data-parallel
environments. A data-parallel task typically depends on multiple input data
blocks. Unless \emph{all} of these blocks are cached in memory, no speedup will result. To capture this \emph{all-or-nothing} property, we propose a
more relevant metric, called \emph{effective cache hit ratio}.
Specifically, a cache hit of a data block is said to be \emph{effective} if it can speed up a compute task. In order to optimize the effective cache hit ratio, we propose the Least
Effective Reference Count (LERC) policy that persists the dependent blocks of
a compute task \emph{as a whole} in memory. We have implemented the LERC policy as a memory manager in Spark and evaluated its performance through Amazon EC2 deployment. Evaluation results demonstrate that LERC helps speed up data-parallel jobs by up to $37\%$ compared with the
widely employed least-recently-used (LRU) policy.

\end{abstract}

\section{Introduction}
\label{sec:intro}

Memory cache plays a pivotal role in data-parallel frameworks, such as Spark \cite{zaharia2012resilient}, Tez \cite{saha2015apache}, Piccolo \cite{power2010piccolo} and Storm \cite{marz2014apachestorm}. By caching input and intermediate data in memory, I/O-intensive jobs can be sped up by orders of magnitude \cite{zaharia2012resilient,ananthanarayanan2012pacman:}. However, compared with stable storage, memory cache remains constrained in production clusters, and it is not possible to persist all data in memory \cite{sharma2016expanded}. Efficient cache management, therefore, becomes highly desirable for parallel data analytics.



\begin{figure}[tbp]
  \centering\includegraphics[width=0.35\textwidth]{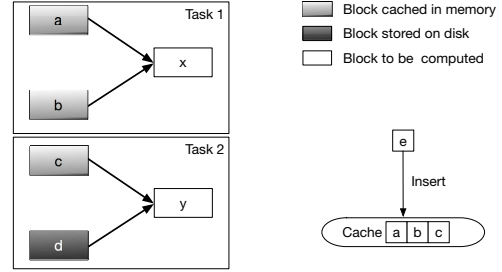}
  \caption{A Spark job with two \texttt{coalesce} tasks. Task 1 (Task 2) coalesces data blocks $a$ and $b$ (data blocks $c$ and $d$) into a larger block $x$ (block $y$). Blocks $a$, $b$ and $c$ are initially persisted in a 3-entry cache, while block $d$ is on disk. Another block $e$ is to be inserted in the cache before Task 1 and Task 2 are scheduled. All the blocks are of unit size.}
  \label{example}
  \vspace{-.4em}
\end{figure}

Unlike caching in storage systems, operating systems and databases, cache management in data-parallel frameworks is characterized by the defining \emph{all-or-nothing} requirement. That is, a compute task cannot be accelerated unless \emph{all} of its dependent datasets, which we call \emph{peers} in this paper, are cached in memory. In Hadoop \cite{shvachko2010hadoop} and Spark \cite{sparkapi}, data-parallel operations, such as \texttt{reduce}, \texttt{join} and \texttt{coalesce}, are typically performed on multiple datasets. Fig.~\ref{example} shows a Spark job consisting of two \texttt{coalesce} tasks. Task 1 (Task 2) fetches two data blocks $a$ and $b$ (blocks $c$ and $d$) and coalesces them into a larger block $x$ (block $y$). In this example, blocks $a$ and $b$ (blocks $c$ and $d$) are \emph{peers} of each other. Caching only one peer provides no benefit, as the task computation is bottlenecked by the reading of the other peer from disk.


 
However, existing cache management policies are \emph{agnostic} to this all-%
or-nothing cache requirement. Instead, they simply optimize the block
\emph{cache hit ratio}, regardless of whether a cache hit can
\emph{effectively} speed up a compute task. For instance, the popular \emph
{least-recently-used} (LRU) policy employed by prevalent parallel frameworks
\cite{zaharia2012resilient,marz2014apachestorm,shinnar2012m3r:,saha2015apache}
caches the data blocks that have been recently used, counting on their future
access to optimize the cache hit ratio. To illustrate that LRU may violate the
all-or-nothing cache requirement for data-parallel tasks, we refer back to the
example in Fig.~\ref{example}. We assume that four unit-sized blocks $a$, $b$,
$c$ and $d$ have been materialized and will be used by two \texttt{coalesce}
tasks to compute blocks $x$ and $y$. The cache can persist 3 blocks and
initially holds blocks $a$, $b$ and $c$. Suppose that another block $e$ will
be inserted to the cache, forcing one in-memory block to be evicted. Among
the three cached blocks, block $c$ is the \emph{only} right choice of
eviction, as caching it alone without its peer $d$ speeds up no task. However,
with the LRU policy, block $c$ will remain in memory unless it is the least
recently used. We shall show in Sec.~\ref{sec:background} that the recently
proposed Least Reference Count (LRC) policy \cite{yinghaolrc} for data-%
parallel frameworks may run into a similar problem.

The all-or-nothing property calls for \emph{coordinated} cache management in
data-parallel clusters. In this work, we argue that cache policy should optimize a more relevant metric, which we call the
\emph{effective cache hit ratio}. In particular, we say a cache hit of a block is \emph{effective}
if it helps speed up a compute task, i.e., all the other peers of this block
are in memory. Referring back to the previous example, caching block $c$
without block $d$ is ineffective as it speeds up no task. Intuitively, the
effective cache hit ratio measures to what degree the all-or-nothing property
can be retained and how likely compute tasks can get accelerated by having all
their dependent datasets kept in memory.


In order to optimize the effective cache hit ratio, we design a coordinated cache management policy, called Least Effective Reference Count (LERC). LERC builds on our previously proposed LRC policy \cite{yinghaolrc} yet retains the \emph{all-or-nothing} property of data-parallel tasks. LERC always evicts the data
block with the smallest \emph{effective reference count}. The effective
reference count is defined, for each data block $b$, as the number of
\emph{unmaterialized} blocks whose computation depends on block $b$ and
\emph{can be sped up by caching}---meaning, the input datasets of the
computation are all in memory.




We have implemented LERC in Spark and evaluated its performance in a 20-node Amazon EC2 cluster against representative workloads. Our prototype evaluation shows that LERC outperforms both LRU and LRC, reducing
the job completion time by $37.0\%$ and $18.6\%$, respectively. In addition, we have confirmed through experiments that compared to the widely adopted cache hit ratio, the proposed effective cache ratio serves as a more relevant metric to measure cache performance in data-parallel systems.


\section{Inefficiency of Existing Cache Policies}
\label{sec:background}

In this section, we introduce the background information and motivate the need for coordinated cache management in data-parallel systems.

  
\subsection{Recency- and Frequency-Based Cache Replacement Policies}

Traditional cache management policies optimize the cache hit ratio by evicting data blocks
based on their access recency and/or frequency. LRU \cite{mattson1970evaluation} and LFU \cite{aho1971principles} are the two representative algorithms.

\begin{itemize}
  \item \textbf{Least Recently Used (LRU)}: The LRU policy always evicts the data block that has not been accessed for the longest period of time. LRU bets on the \emph{short-term} data popularity. That is, the recently accessed data is assumed to be likely used again in the near future. LRU is the default cache replacement policy in many popular parallel frameworks, such as Spark \cite{zaharia2012resilient}, Tez \cite{saha2015apache} and Storm \cite{marz2014apachestorm}.

  \item \textbf{Least Frequently Used (LFU)}: The LFU policy always evicts the data block that has
  been accessed the least times. Unlike LRU, LFU bets on the \emph{long-term} data popularity. That is, the data accessed frequently in the past will likely remain popular in the future. 
\end{itemize}

Recency and frequency can also be used in combination, e.g., LRFU
\cite{lee2001lrfu} and K-LRU \cite{o1993lru}. All of these cache algorithms
predict future data access based on historical information, and are typically
employed in operating systems, storage, database and web servers where the
underlying data access pattern cannot be known \emph{a priori}.

\subsection{Dependency-Aware Cache Management}

In data-parallel frameworks such as Spark \cite{zaharia2012resilient} and Tez
\cite{saha2015apache}, compute jobs have rich semantics of data dependency
in the form of directed acyclic graphs (DAGs). These dependency DAGs dictate the
underlying data access patterns, providing new opportunities
for dependency-aware cache management.

\begin{figure}[tbp]
  \centering\includegraphics[width=0.24\textwidth]{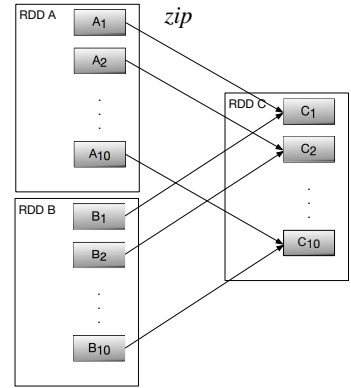}
  \caption{DAG of the \texttt{zip} job in Spark with three RDDs $A$, $B$ and $C$. Each block in RDD $C$ depends on two corresponding blocks from RDD $A$ and RDD $B$.} 
  \label{DAG}
  \vspace{-.4em}
\end{figure}

For example, Fig.~\ref{DAG} shows the DAG of a Spark \texttt{zip} job
\cite{sparkapi}. Spark manages data through an abstraction called Resilient
Distributed Datasets (RDDs) \cite{zaharia2012resilient}. Each RDD is
partitioned into multiple blocks across different machines. In Fig.~\ref{DAG},
there are three RDDs $A$, $B$ and $C$, each consisting of $10$ blocks. RDD $C$
is a collection of key-value pairs obtained by zipping RDD $A$ with RDD $B$.
The computation of each block $C_i$ depends on two blocks $A_i$ (key) and $B_i$ (value).

In Spark, the data dependency DAG is readily available to
\texttt{DAGScheduler} upon a job submission. The recently proposed Least
Reference Count (LRC) policy \cite{yinghaolrc} takes advantage of this DAG
information to determine which RDD block should be kept in memory.



\noindent \textbf{Least Reference Count (LRC)}: The LRC policy always evicts the data block with the least \emph{reference count}. The reference count is defined, for each block, as the number of unmaterialized blocks depending on it. In the example of Fig.~\ref{DAG}, each block of RDD $A$ and RDD $B$ has reference count 1. 

Intuitively, the higher the reference count an RDD block has, the more to-be-scheduled tasks depend on it, and the higher probability that this block will be used in the near future.
In many applications, some RDD blocks would be used iteratively during the computation with much higher reference counts than others, e.g., the training datasets for cross-validation in machine learning \cite{kohavi1995study}. With LRC, these blocks would have a higher chance to be cached in memory than others.

\subsection{All-or-Nothing Cache Requirement}
\label{sec:all-or-nothing}

Prevalent cache algorithms, be it recency/frequency-based or dependency-%
aware, settle on the cache hit ratio as their optimization objective. However,
the cache hit ratio fails to capture the \emph{all-or-nothing} 
requirement of data-parallel tasks and is not directly linked to their
computation performance. The computation of a data-parallel task usually
depends on multiple data blocks, e.g., \texttt{join}, \texttt{coalesce} and
\texttt{zip} in Spark \cite{sparkapi}. A task cannot be sped up unless
\emph{all} its dependent blocks, which we call \emph{peers} (e.g., block $A_1$
and block $B_1$ in Fig.~\ref{DAG}), are cached in memory.

\begin{figure}[tbp]
  \centering\includegraphics[width=0.3\textwidth]{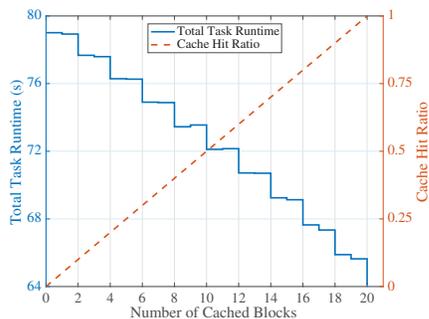}
  \caption{Total task runtime of the example job in Fig.~\ref{DAG}, with the cached RDD blocks increasing one at a time in the order of $A_1$, $B_1$, $A_2$, $B_2$, $\dots$, $A_{10}$, $B_{10}$.} 
  \label{StickyExp}
  \vspace{-.1in}
\end{figure}

\vspace{.4em}
\noindent\textbf{Measurement study:}
To demonstrate the all-or-nothing property in data-parallel systems, we ran the 
Spark \texttt{zip} job in an Amazon EC2 cluster with 10
\texttt{m4.large} instances \cite{ec2}. The job DAG is illustrated in Fig.~\ref{DAG},
where each of the two RDDs $A$ and $B$ is configured as 200 MB. We repeatedly
run the \texttt{zip} job in rounds. In the first round, no data block is cached
in memory. In each of the subsequent rounds, we 
add one more block to cache, following the caching order $A_1, B_1, A_2, B_2, \dots, A_{10}, B_{10}$.
Eventually, all 20 blocks are cached in memory in the final round. In each round, we measure
the cache hit ratio and the total runtime of all 10 tasks. Fig.~\ref{StickyExp} depicts
our measurement results against the number of RDD blocks in memory.
Despite the linearly growing cache hit ratio with more in-memory blocks, the
task completion time is notably reduced only after the two peering blocks $A_i$ and $B_i$ have been cached. 


\vspace{.4em}
\noindent\textbf{Inefficiency of existing cache policies:}
The first step to meet the all-or-nothing cache requirement is to identify
which blocks are peers with each other and should be cached together as a
whole. This information can only be learned from the data dependency DAG, but
it has not been well explored in the literature. Many existing cache policies,
such as LRU and LFU, are \emph{oblivious} to the DAG information, and are
unable to retain the all-or-nothing property. The recently proposed LRC
\cite{yinghaolrc} policy, though DAG-aware, does not differentiate the peering
blocks and hence suffer from the same inefficiency as LRU. By referring back to the
previous example in Fig.~\ref{example}, we see that blocks $a$, $b$ and $c$
have the same reference count of 1 and would have an equal chance
to get evicted by LRC. In other words, LRC would evict a wrong block (other
than $c$) with probability $67\%$. To the best of our knowledge, PACMan
\cite{ananthanarayanan2012pacman:} is the only work that tries to meet the
all-or-nothing requirement for cache management in parallel clusters. However,
PACMan is \emph{agnostic} to the semantics of job DAGs, and its objective is to speed
up data sharing across different jobs by caching \emph{complete} datasets (HDFS files).
Since PACMan only retains the all-or-nothing property for each individual dataset,
if a job depends on multiple datasets, completely caching a subset of them
provides no performance benefits.


\vspace{.4em}
In summary, we have shown through a toy example and measurement study that optimizing the cache
hit ratio---the conventional performance metric employed by existing cache
algorithms---does not necessarily speed up data-parallel computation
owing to its all-or-nothing cache requirement. 
We shall design a new cache management policy that meets this
requirement based on the peer
information extracted from the data dependency DAGs.




\section{Least Effective Reference Count}
\label{sec:lerc}

In this section, we define the \emph{effective cache hit ratio} as a more relevant
cache performance metric for data-parallel tasks. To optimize this new metric,
we propose a coordinated cache management policy, called Least
Effective Reference Count (LERC). We also present our implementation of LERC
as a cache manager in Spark.

\subsection{Effective Cache Hit Ratio}

In a nutshell, a cache hit of a block is \emph{effective} if it can speed up the
computation. In data-parallel frameworks, a compute task typically depends on
multiple data blocks. We call all these blocks \emph{peers} with respect to
(w.r.t.) this task. Caching only a subset of peering blocks provides no speedup
for the task. Formally, we have the following definition.


\begin{definition}[Effective Cache Hit]
  For a running task, the cache hit of a dependent block is \emph{effective} if 
  its peers w.r.t. the task are \emph{all} in memory.
\end{definition}

We now define the \emph{effective cache hit ratio} as the number of effective cache
hits normalized by the total number of block accesses. This ratio
directly measures how much compute tasks can benefit from data caching. 

By referring back to the previous example in Fig.~\ref{example}, each of the four
blocks $a$, $b$, $c$ and $d$ is used only once in the computation. Initially,
blocks $a$, $b$ and $c$ are in memory. Since block $d$ is on disk, evicting
its peer $c$ has no impact on the effective cache hit ratio, which remains at
$50\%$ (i.e., two effective cache hits for $a, b$ out of 4 block accesses).
Evicting block $a$ (or $b$), on the other hand, results in no effective cache
hit. 

The effective cache hit ratio directly measures the performance of a cache
algorithm. Algorithms that optimize the traditional cache hit ratio may not
perform well in terms of this new metric. Back to the previous example, the
LRC policy \cite{yinghaolrc} evicts each of the three in-memory blocks, $a$,
$b$ and $c$, with an equal probability (cf. Sec.~\ref{sec:all-or-nothing}),
leading to a low effective cache hit ratio of $\frac{1}{3} \times 50\% +
\frac{2}{3} \times 0\% = 16.7\%$ in expectation. The LRU policy can be even worse as the effective hit ratio is 0, unless block $c$ is the least recently used. 


A naive approach to optimize the effective cache hit ratio is the \emph{sticky
eviction} policy. That is, the peering data blocks stick to each other and are
evicted as a whole if any of them is not in memory. However, such a sticky
policy can be highly inefficient. A data block might be a shared input of
multiple tasks. Caching the block, even though not helping speed up one task
(i.e., not all peers w.r.t. the task are in memory), may benefit another. With
the sticky policy, this block would surely be evicted, and no task can be sped up.
We propose our solution in the next subsection.





\subsection{Least Effective Reference Count Caching}



We start with the definition of the \emph{effective reference count}. In data-parallel
frameworks, a block may be referenced by many tasks (i.e., used as input). We differentiate \emph{effective reference}
from the others by the following definition. 
\begin{definition}[Effective Reference]
  Let block $b$ be referenced by task $t$. 
  We say this reference is \emph{effective} if task $t$'s dependent blocks, if \emph{computed}, are all cached in memory.
\end{definition}

For a data block, the \emph{effective reference count} is simply the number of
its effective references, which, intuitively, measures how many downstream
tasks can be sped up by caching this block. Based on this definition, we
present the \emph{Least Effective Reference Count} (LERC) policy as follows.

\vspace{.4em}
\noindent \textbf{Least Effective Reference Count (LERC):}
\emph{The LERC policy evicts the block whose effective reference count is the smallest.}



\vspace{.4em}
As a running example, we refer back to Fig.~\ref{example}. Each of blocks $a$ and $b$ has effective reference count 1, while block $c$ has no effective reference count (its reference
by Task 2 is not effective as block $d$ is not in memory). With the LERC policy, block $c$
is evicted, which is the optimal decision in this example.

LERC has two desirable properties. First, by prioritizing blocks with large
effective reference counts, LERC is able to retain the peering blocks as
entities as much as possible, through which the effective cache hit ratio is
optimized. Second, the effective reference count can be easily extracted from
the job DAG that is readily available to the scheduler. 



\subsection{Spark Implementation}

Implementing LERC in data-parallel frameworks, such as Spark, poses some non-%
trivial challenges. Maintaining a block's effective reference count requires
knowing the caching status of its peers on different workers. Synchronizing
each block's caching status across workers may incur significant communication
overhead. The first approach is to maintain the caching status profile in a
centralized manner at the Spark driver. When a block is evicted from memory,
the worker reports to the driver. The driver updates the effective reference
counts of other peering blocks and broadcasts the updates to all workers.
Alternatively, we can let workers maintain the caching status in a distributed
manner. Upon a block eviction on a worker, all the other workers must be
notified immediately to update the effective reference counts. Both approaches
broadcast a large amount of information across the network, and are expensive
to implement.

We next present our implementation to address this problem.


\begin{figure}[tbp]{}
\centering\includegraphics[width=7cm]{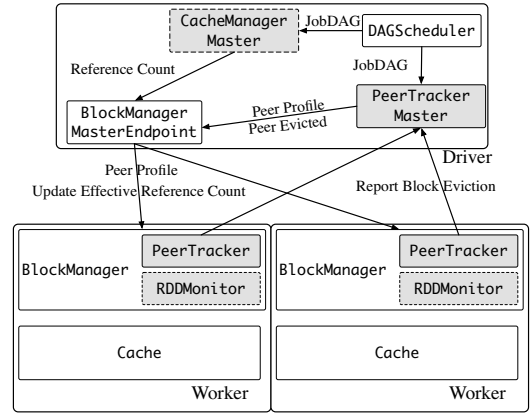}
\caption{Overall system architecture of the proposed coordinated cache manager in Spark. Our implementation modules are highlighted as shaded boxes with solid borderlines. Shaded boxes with dashed borderlines are the legacy implementations of the LRC work \cite{yinghaolrc}.
}\label{architecture}
\vspace{-.1in}
\end{figure}

\vspace{.4em}
\noindent \textbf{Architecture overview.} Fig.~\ref{architecture} presents an architecture overview of our LERC cache manager in Spark. The \texttt{CacheManagerMaster} and \texttt{RDDMonitor} are the legacy modules of our LRC implementation \cite{yinghaolrc}. The \texttt{CacheManagerMaster} in the driver parses the reference count profile and maintains it together with the \texttt{RDDMonitor}s in workers. In this paper, we have implemented two components: (1) \texttt{PeerTrackerMaster} that profiles the peer information in the job DAG obtained from the \texttt{DAGScheduler}, and (2) \texttt{PeerTracker} in each worker that reports the status of the peer blocks when necessary. 

\vspace{.4em}
\noindent \textbf{Workflow.} When the Spark driver is launched, the \texttt{PeerTrackerMaster} is initialized together with other components in the driver. Upon a job submission, the \texttt{PeerTrackerMaster} obtains the job DAG from the \texttt{DAGScheduler} and parses out the peer information. The peer information profile is then broadcast via the \texttt{BlockManagerMasterEndpoint} to all \texttt{PeerTracker}s in the cluster. The \texttt{PeerTracker} initially labels every \emph{peer-group} (all peering blocks w.r.t. a task) as ``complete,'' and then changes it to ``incomplete'' if any of its materialized block has been evicted from memory. Upon a block eviction, the \texttt{PeerTracker} first checks whether this block belongs to any ``complete'' peer-group. If so, the effective reference counts of blocks in these peer-groups should be updated due to this block eviction. A block eviction report is sent to the \texttt{PeerTrackerMaster} and broadcast to other workers. Upon receiving a block eviction message, the \texttt{PeerTracker} scans all the ``complete'' peer-groups. If there are ``complete'' peer-groups containing the evicted block, the \texttt{PeerTracker} labels them as ``incomplete''  and informs the \texttt{RDDMonitor} to decrease the effective reference counts of the peering blocks accordingly.

\vspace{.4em}

\noindent \textbf{Communication Overhead.} Our implementation minimizes the number of required communication messages. To see this, we first show that at most one broadcasting is triggered for the entire group of peer blocks in our implementation. By labeling the ``complete'' peer-groups locally in the \texttt{PeerTracker}s, there is no need to track the caching status of each peer block separately. Only the block eviction in a ``complete'' peer-group triggers the updating of the peering blocks' effective reference counts. Once a block eviction message is broadcast, the peer-group becomes ``incomplete'', and no more updating messages will be required for this peer-group. We next show that it is necessary to broadcast at least one block eviction message for each group, if any of its peer has been evicted. Since it is possible that some of the evicted block's peers have not been computed yet, the block eviction message should be broadcast to all workers instead of those with the evicted block's peers. 




\section{Evaluation}
\label{sec:evaluation}

In this section, we evaluate the efficacy of the LERC cache manager with synthesized workloads in Amazon EC2. We also confirm that the effective cache hit ratio is a more relevant metric of cache performance than the widely used cache hit ratio. 

\vspace{.4em}
\noindent \textbf{Cluster deployment.}
Our implementation is based on Spark 1.6.1. In order to highlight the
advantage of memory locality, we disabled the OS page cache by triggering direct disk I/O from/to the hard disk. We deployed an Amazon EC2 \cite{ec2} cluster with 20 nodes of type \texttt{m4.large}, each with a dual-core 2.4 GHz Intel Xeon E5-2676 v3 (Haswell) processor and 8 GB memory.

\vspace{.4em}
\noindent \textbf{Experiment settings.} In the experiment, we simulated $10$ tenants submitting Spark \texttt{zip} \cite{sparkapi} jobs in parallel. In each job, two files of $400$ MB are firstly partitioned into $100$ blocks and stored in the $20$ machines. Since the 10 jobs operate on different files, the total size of the input blocks is $400 \text{ MB} \times 2 \times 10 = 8 \text{ GB}.$ The cache manager will decide which blocks should be stored on disk when the cache is full. After that, $100$ zip tasks are scheduled for each job to zip the two files into $100$ key-value pairs, where the keys are the data of the first file, and the values are the data of the second file. Notice that only when both the key and value are cached in memory will a \texttt{zip} task be sped up. We measure the total completion time of the $10$ jobs to compare the performance of different cache policies.

\subsection{Job Completion Time}
\begin{figure}[htbp]{}
  \centering\includegraphics[width=0.38\textwidth]{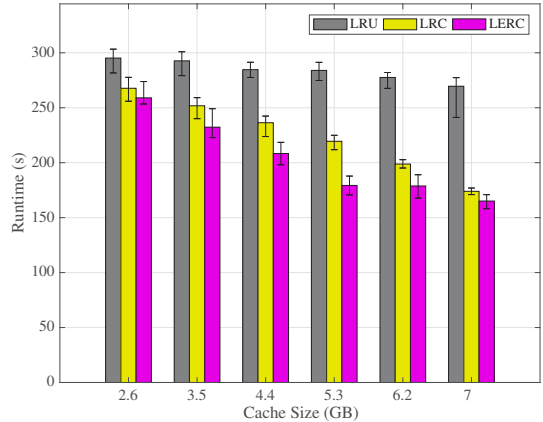}
  \caption{Experiment runtime under the three cache management policies with different cache sizes.}
  \label{runtime}
\end{figure}

We conducted the experiment using three cache replacement policies, i.e., LRU, LRC, and LERC, with different memory cache sizes. In particular, we configured \texttt{storage.memoryFraction} in the legacy Spark to throttle the memory used for RDD caching to a given size. We measured the total experiment runtime, i.e., the make span of the $10$ submitted jobs. We repeated each experiment $10$ times and depict the average results in Fig.~\ref{runtime}. The error bars show the maximum and minimum completion time in the 10 runs.

To our expectation, as the size of RDD cache increases, the total experiment runtime decreases under all the three cache policies. In all cases, LRC consistently outperforms the default LRU policy. LERC further reduces the experiment completion time over LRC. When the cache size is $5.3$ GB, for instance, the average runtimes under the three policies are  $284$ s (LRU), $220$ s (LRC) and $179$ s (LERC), respectively. The LERC policy speeds up job completion by $37.0\%$ and $18.6\%$ compared to the LRU and LRC policies, respectively. 

\subsection{Effective Cache Hit Ratio}
\begin{figure}[htbp]{}
  \centering\includegraphics[width=0.38\textwidth]{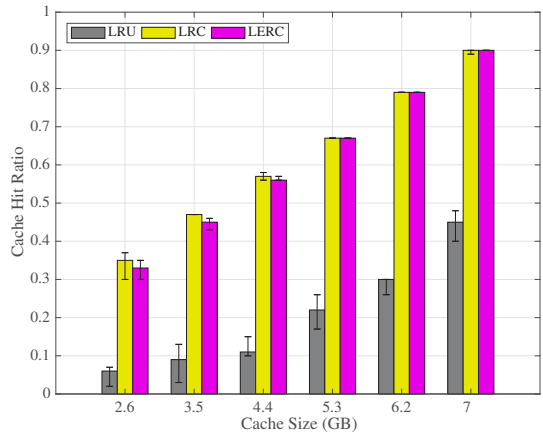}
  \caption{Cache hit ratio under the three cache management policies with different cache sizes.}
  \label{cachehitratio}
  \vspace{-.1in}
\end{figure}

\begin{figure}[htbp]{}
  \centering\includegraphics[width=0.38\textwidth]{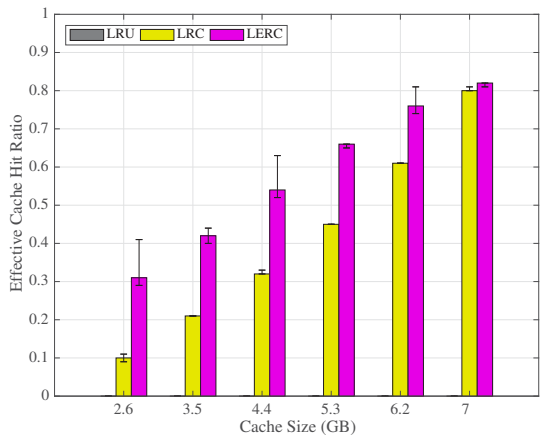}
  \caption{Effective cache hit ratio under the three cache management policies with different cache sizes. Notice that the effective cache hit ratio is near zero under the LRU policy.}
  \label{effectivecachehitratio}
  \vspace{-.1in}
\end{figure}

We now evaluate the relevance of the two metrics, i.e., the effective cache hit ratio and the cache hit ratio, in measuring the cache performance in data-parallel clusters. Both of the two metrics were recorded in the previous experiments. The results are shown in Fig.~\ref{cachehitratio} and Fig.~\ref{effectivecachehitratio}. 

Fig.~\ref{cachehitratio} shows that LRC achieves the highest cache hit ratio, while LERC closely follows. This is because LRC aims to optimize the cache hit ratio, and it outperforms LRU by taking
advantage of the DAG information. LERC also takes use of this information, but it gives up on retaining those ineffective cache hits that are unable to speed up tasks. It is for this reason that the cache hit ratio is slightly compromised.

Fig.~\ref{effectivecachehitratio} shows that LERC always achieves the highest effective cache hit ratio. The smaller the cache size is, the more advantageous LERC is. We therefore conclude that LERC is able to make the best use of the constrained cache resources to effectively speed up compute tasks. As the available cache space increases, LRC is more likely to retain an entire peer-group in memory, and its effective cache hit ratio becomes closer to that of LERC. On the other hand, when the cache size is small, many peer blocks will have to be evicted out of memory. Since LERC needs to maintain the effective reference count, a salient communication overhead is incurred.
For this reason, in Fig.~\ref{runtime}, LERC does not save much job runtime compared with LRC when the cache volume is relatively small, even though it achieves a much higher effective cache hit ratio. As the cache size increases, less communication cost is incurred, and the performance in the effective cache ratio gets in line with the job runtime.

Notice that the effective cache hit ratio of LRU is always near zero in our experiment. Since the tenants submit their jobs in parallel, the first file (keys required by \texttt{zip}) of each job is highly likely to be replaced by the second file (values required by \texttt{zip}) of other jobs arriving \emph{later} under the LRU policy. Therefore, when the \texttt{zip} starts, only the values are cached, resulting in zero effective cache hit.

\vspace{.4em}
We conclude this section by noting that although LRC achieves the highest cache hit ratio throughout the experiments, it incurs longer job completion time compared with LERC. On the other hand, LERC achieves a higher effective cache hit ratio over LRC. Its relative advantage in the effective cache hit ratio is consistent with that in the job completion time (when the communication overhead is not a concern). We therefore draw the conclusion that the effective cache hit ratio serves as a more relevant metric for cache performance in our experiments.

\section{Conclusions}
\label{sec:conclusion}

In this paper, we have identified the defining all-or-nothing property in data-parallel systems, i.e., a compute task can only be sped up when \emph{all} of its input datasets are cached in memory. Existing cache management policies are agnostic to this all-or-nothing requirement. Instead, they settle on the cache hit ratio as their optimization objective and hence
cannot effectively reduce the task runtime in data-parallel environments. To address this problem, we have proposed the effective cache hit ratio as a more relevant cache performance metric for data-parallel tasks. We have designed a coordinated cache policy, Least Effective Reference Count (LERC), that optimizes this metric by evicting the data blocks with the smallest effective reference count. We have implemented LERC as a pluggable cache manager in Spark, and evaluated its performance through Amazon EC2 deployments. Experimental results validated the relevance of the effective cache hit ratio and the performance advantage of the LERC policy. Compared with the popular LRU policy and the recently proposed LRC policy, LERC speeds up the job completion by up to $37\%$ and $19\%$, respectively.



\bibliography{main}
\bibliographystyle{IEEEtran}

\end{document}